\documentclass[aps,prd,twocolumn,showpacs,superscriptaddress,groupedaddress]{revtex4-1} 

\usepackage{graphicx}  
\usepackage{amssymb}             
\usepackage{dcolumn}   
\usepackage{bm}        
\usepackage{amssymb}   
\usepackage[namelimits]{amsmath}
\usepackage{subfigure}
\usepackage{verbatim} 
\usepackage{multirow}

\newcommand{\Eqref}[1]{Eq.~\eqref{#1}}
\newcommand{\EEqref}[1]{Equation~\eqref{#1}}

\newcommand{\Fgref}[1]{Figure~\ref{#1}}
\newcommand{\fgref}[1]{Fig.~\ref{#1}}

\usepackage[colorlinks,linkcolor=red,anchorcolor=blue,citecolor=green]{hyperref}
\usepackage{url}
\usepackage[]{lineno}
\usepackage{cleveref}

\begin{document}
\title{Measurement of transverse single-spin asymmetries of $\pi^0$ and electromagnetic jets at forward rapidity in 200 and 500~GeV  transversely polarized proton-proton collisions}

\date{\today}
\author{
J.~Adam$^{6}$,
L.~Adamczyk$^{2}$,
J.~R.~Adams$^{39}$,
J.~K.~Adkins$^{30}$,
G.~Agakishiev$^{28}$,
M.~M.~Aggarwal$^{41}$,
Z.~Ahammed$^{61}$,
I.~Alekseev$^{3,35}$,
D.~M.~Anderson$^{55}$,
A.~Aparin$^{28}$
% ...
}

% ...
\address{$^{2}$AGH University of Science and Technology, FPACS, Cracow 30-059, Poland}
\address{$^{3}$Alikhanov Institute for Theoretical and Experimental Physics NRC "Kurchatov Institute", Moscow 117218, Russia}
% ...
\address{$^{6}$Brookhaven National Laboratory, Upton, New York 11973}
% ...
\address{$^{28}$University of Jammu, Jammu 180001, India}
% ...
\address{$^{30}$Kent State University, Kent, Ohio 44242}
% ...
\address{$^{35}$Michigan State University, East Lansing, Michigan 48824}
% ...
\address{$^{39}$Nuclear Physics Institute of the CAS, Rez 250 68, Czech Republic}
% ...
\address{$^{41}$Institute of Nuclear Physics PAN, Cracow 31-342, Poland}
% ...
\address{$^{55}$Temple University, Philadelphia, Pennsylvania 19122}
% ...
\address{$^{61}$Valparaiso University, Valparaiso, Indiana 46383}
% ...

\collaboration{STAR Collaboration}\noaffiliation

\begin{abstract}
	The STAR Collaboration reports measurements of the transverse single-spin asymmetry (TSSA) of inclusive $\pi^0$ at center-of-mass energies ($\sqrt s$) of 200~GeV  and 500~GeV  in transversely polarized proton-proton collisions in the pseudo-rapidity region 2.7 to 4.0. 
	The results at the two different energies show a continuous increase of the TSSA with Feynman-$x$, and, when compared to previous measurements, no dependence on $\sqrt s$ from 19.4~GeV  to 500~GeV  is found. To investigate the underlying physics leading to this large TSSA, different topologies have been studied. $\pi^0$ with no nearby particles tend to have a higher TSSA than inclusive $\pi^0$. The TSSA for inclusive electromagnetic jets, sensitive to the Sivers effect in the initial state, is substantially smaller, but shows the same behavior as the inclusive $\pi^0$ asymmetry as a function of Feynman-$x$. To investigate final-state effects, the Collins asymmetry of $\pi^0$ inside electromagnetic jets has been measured. The Collins asymmetry is analyzed for its dependence on the $\pi^0$ momentum transverse to the jet thrust axis and its dependence on the fraction of jet energy carried by the $\pi^0$. The asymmetry was found to be small in each case for both center-of-mass energies. 
	All the measurements are compared to QCD-based theoretical calculations for transverse-momentum-dependent parton distribution functions and fragmentation functions. Some discrepancies are found, which indicates new mechanisms might be involved.
\end{abstract}

\maketitle

\section{introduction}
Significant transverse single-spin asymmetries (TSSA) have been observed for charged and neutral-hadron production in hadron-hadron collisions over a wide range of collision energies since the 1970’s~\cite{AN1977,E7042,2008AN,PHENIX,Rhicf}.
The early leading-order QCD calculation showed the corresponding asymmetry is exceedingly small~\cite{Kane:1978nd}.
Different models and mechanisms have been proposed to understand these sizable asymmetries~\cite{Liang:2000gz,DAlesio:2007bjf,Chen:2015tca}. Recently, all of the QCD-based formalisms for TSSA have been categorized into two frameworks. The first one is based on transverse-momentum-dependent (TMD) parton distribution or fragmentation functions, and the second one is based on Twist-3 collinear factorization. These two ans\"{a}tze probe different underlying sub-processes. In the TMD framework one requires two scales, a large momentum transfer $Q$ as a ``hard" scale, and a modest transverse momentum $q_\mathrm{T}$, as a ``soft" scale. In general one requires $Q \gg q_\mathrm{T}$. Calculations in the Twist-3 framework only require one scale with $q_\mathrm{T} \gg \Lambda_\mathrm{QCD}$, the strong interaction scale. It has been proven~\cite{TMDunite} that both approaches describe the same physics in the kinematic region where they overlap, i.e., $Q \gg q_\mathrm{T} \gg \Lambda_\mathrm{QCD}$.   

For both frameworks, the origin of the hadron TSSA in hadron-hadron collisions can have two sources, namely an initial and a final-state effect. In the pure TMD approach, the initial-state effect is from the Sivers function ($f^{\perp,q}_\mathrm{1T}$)~\cite{Sivers}, and the final-state effect is from the coupling of the chiral-odd transversity parton distribution function (PDF) and the chiral-odd Collins fragmentation function ($H_1^\perp$)~\cite{Collins:1992kk,Collins:1993kq}. The counterpart of the Sivers function in Twist-3 collinear factorization is the Efremov-Teryaev-Qui-Sterman (ETQS) function ($T_{q,F}$)~\cite{Efremov:1981sh,Qiu:1991pp}. It has been shown that $T_{q,F}$ is related to the Sivers function~\cite{Boer:2003cm} through the following relation:
\begin{equation}\label{eq1}
    T_{q,F}(x,x)=-\int {d^2k_{\perp}\frac{|k^2_{\perp}|}{M}f_{1T}^{\perp,q}(x,k^2_{\perp})|_\mathrm{SIDIS}}.
\end{equation}
Therefore the Sivers function extracted from semi-inclusive deep-inelastic scattering (SIDIS) data can be used to constrain the ETQS-function in transversely polarized proton-proton collisions. A very similar relation holds for the Collins fragmentation function equivalent in the Twist-3 formalism~\cite{Yuan:2009dw}.

In the measurements discussed in this paper, the large transverse momentum ($p_\mathrm{T}$) of the final-state $\pi^0$ fits the scale requirement of the Twist-3 formalism. Many phenomenological studies of the pion TSSA have been done in the Twist-3 framework. The contributions from initial-state effects~\cite{Boros:1993ps,twist3-I1,twist3-I2,twist3-I5}, final-state effects~\cite{Yuan:2009dw,Kang:2010zzb,twist3-F1,twist3-F2} and their combination~\cite{twist3-1,Zhou:2017sdx} have been calculated. 
For many years the initial-state effect was thought to be the main source of the TSSA. However, it has been realized that the ETQS-function extracted from proton-proton collisions and the Sivers function extracted from SIDIS do not coincide well~\cite{twist3-1}. In recent years, it was proposed that the initial-state effects are small and the final-state effects are the main contribution to the TSSA~\cite{Zhou:2017sdx,Twist3Final}.  

The initial-state and final-state effects cannot be disentangled for the pion TSSA, but other observables such as the jet TSSA and Collins asymmetry can be used to separate them. The TSSA for jets is considered to be sensitive to initial-state effects. An earlier measurement in transversely polarized proton-proton collisions at 
$\sqrt{s}$ = 500~GeV  by the $A_\mathrm{N}DY$ experiment found the inclusive jet TSSA to be very small~\cite{ANDY}. This was reproduced by theoretical calculations~\cite{jetAN1,jetAN2} of the jet TSSA. On the other hand, the Collins asymmetry is only sensitive to final-state effects. It measures the azimuthal asymmetry of a hadron within a jet originating from the fragmentation of a transversely polarized quark. Theory predictions for the Collins asymmetry in transversely polarized proton-proton collisions can be found in Refs.~\cite{Kang:2017btw,fengCollins,DAlesio:2017bvu}. Experimental results at mid-rapidity have been reported by the STAR (Solenoidal Tracker At RHIC) Collaboration~\cite{STARCollins}. 

In this paper, the STAR Collaboration at the Relativistic Heavy Ion Collider (RHIC) reports new measurements of the TSSA for the inclusive $\pi^0$ production at large rapidity in transversely polarized proton-proton collisions at $\sqrt s$ of 200 and 500~GeV  to study the energy dependence of the TSSA. To understand the underlying physics mechanisms, different topologies for the TSSA have been investigated, which include the
extraction of the TSSA for inclusive and isolated $\pi^0$, electromagnetic jets, and the Collins effect through $\pi^0$ inside an electromagnetic jet. 
Recently STAR published a complementary study of the nuclear dependence of the $\pi^0$ TSSA~\cite{nucl_dep_paper}, which used the same 200~GeV  proton-proton data that are investigated here.  Although some technical aspects of the two analyses slightly differ, the results are consistent in those cases where the same quantity is measured. 

This paper is organized as the following.  Sec.~II provides the analysis details including a brief overview of RHIC and the Forward Meson Spectrometer (FMS) detector, event selection, $\pi^0$ and jet reconstruction, and the methods of spin asymmetry calculation. The correction and systematic uncertainty studies are discussed in Sec.~III. Section IV gives the TSSA results for inclusive $\pi^0$, isolated $\pi^0$ and jets, and the Collins asymmetry results for $\pi^0$. Finally, Sec.~V presents a summary of the measurements.

\section{Analysis}
\subsection{Experiment}
The measurements have been performed with the STAR detector~\cite{STAR} at RHIC located at Brookhaven National Laboratory. RHIC is currently the only facility in the world that can provide high energy, high luminosity, highly polarized proton-proton collisions. The clockwise and counterclockwise proton beams at RHIC are labeled as blue and yellow, respectively.
The beam polarization measurements are provided by the RHIC polarimeter group, which develops, maintains and operates the RHIC polarimeters. The details of the beam polarization measurements in recent years can be found in Ref.~\cite{spin}.

\begin{figure}[htbp]
	\centering
	\includegraphics[scale = 0.3]{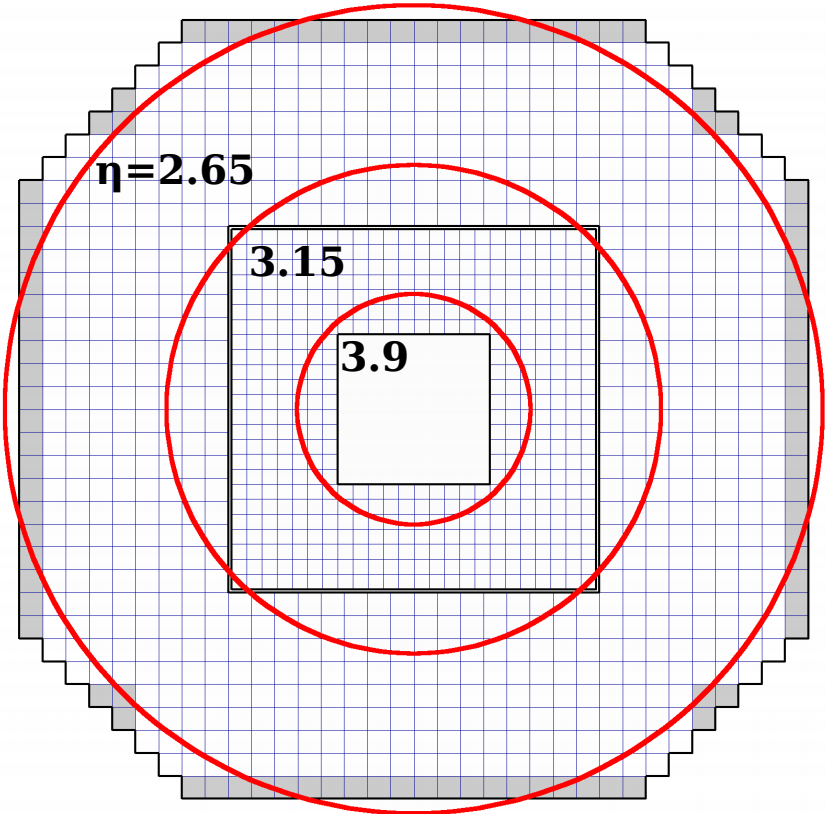}\\
	\caption{The layout of the FMS detector~\cite{ChrisPaper}.
	}\label{fg:FMS}
\end{figure}

The analysis in this paper uses the FMS detector at STAR to reconstruct photons and $\pi^{0}$s. The FMS is an electromagnetic calorimeter installed on the west side of the STAR detector, about 7 meters away from the interaction point. It faces the blue beam with a pseudorapidity coverage of about $2.6 < \eta < 4.1$. 
The layout of the FMS is shown in \fgref{fg:FMS}. The FMS has an octagonal shape with a radius of about 1 m surrounding the beam pipe with a 40~cm $\times$ 40~cm central cutout. The FMS is made of 1264 lead glass towers of two types, which differ in size and density of towers. The towers closer to the beam line are smaller in size in order to separate photons from high energy $\pi^0$ decays. The inner towers are 3.8~cm $\times$ 3.8~cm $\times$ 45~cm in size and cover a pseudorapidity range from 3.3 to 4.1. The outer towers are larger, 5.8~cm $\times$ 5.8~cm $\times$ 60~cm in size, and cover a pseudorapidity range from 2.6 to 3.3.  All towers are wrapped in thin aluminized mylar for optical isolation.  Both tower types have more than 18 radiation lengths, so photons deposit nearly all of their energy in the detector. A detailed description of the detector can be found in Ref.~\cite{Christhesis,ChrisPaper,Braidot:2011zj}.

The calibration of the FMS is based on the invariant mass of the reconstructed $\pi^{0}$. Since the decay photons from the $\pi^{0}$ cover multiple towers, iterations are performed until the gains of all the towers have converged. 

\subsection{Event selection}
The data sets used in this paper were collected by STAR in 2011 and 2015 from transversely polarized proton-proton collisions at $\sqrt s$ of 500 and 200~GeV, respectively. 
The beam polarization of the data set is 52.4 $\pm$ 1.8\,\% for the blue beam, which faces the FMS, in 2011 and 56.6 $\pm$ 1.7\,\% for the blue beam in 2015.

The proton-proton collision events were triggered by the FMS itself, based on the total transverse  energy ($E_\mathrm{T}$) deposited in the detector.
There were two types of triggers used in the analysis, which differ in how the regions for the energy deposition were chosen. The Board Sum triggers were based on the energy sum of overlapping areas, which covered a patch of 4 $\times$ 8 adjacent towers. The Jet Patch triggers used combinations of several non-overlapping Board Sum regions. Each Jet Patch covered a quarter of the FMS with a $\pi$/2 coverage in azimuth. In 2011, there were four patches, with each covering one quadrant.  In 2015, two additional patches straddling the horizontal axis were added to smooth the acceptance. In order to avoid a possible bias from the Board Sum triggers, only Jet Patch triggered data were used in the jet TSSA analysis. 

Each trigger condition was operated with multiple $E_\mathrm{T}$ thresholds. The thresholds for the small-tower Board Sum triggers were 1.6/2.7~GeV  in 2011 and 1.3/1.8/2.2/2.5~GeV  in 2015. For the large-tower Board Sum triggers, the thresholds were 2.7/4.3~GeV  in 2011 and 1.3/1.8/2.8~GeV  in 2015. For the Jet Patch triggers, the thresholds are 2.7/4.3~GeV in 2011 and 1.8/2.8/3.7~GeV  in 2015. The triggers with the lower thresholds were pre-scaled due to the limited bandwidth of the STAR data acquisition system. In the $\pi^0$/EM-jet TSSA analysis, the $\pi^0$/EM-jet $p_\mathrm{T}$ is required to be larger than the trigger threshold of the event.  

The longitudinal vertex position ($z$-vertex) of FMS events was provided by the Beam Beam Counters (BBC) at STAR~\cite{Kiryluk:2005gg}. The $z$-vertex selection for 500 GeV data was $-68$~cm $< z <$ 68~cm, and $-126$~cm $< z <$ 54~cm for 200 GeV data. The latter vertex range was biased towards the negative direction due to the FMS trigger system setup in 2015.     

In 2015, the installation of the Heavy Flavor Tracker (HFT)~\cite{HFT} in STAR introduced some non-collision background, which has an impact on the jet TSSA analysis. In the jet selection, these events are removed effectively with cuts based on information from the BBC and Time-of-Flight (TOF)~\cite{TOF} sub-detectors. For the east BBC, which covers the pseudorapidity range $-5 <\eta < -3.2$ on the opposite side of STAR from the FMS, it was required that at least one tile fired. For the TOF, which covers the mid-rapidity region  $-0.9< \eta < 0.9$, its multiplicity was required to be greater than two.  The non-collision background was found to much less affect the $\pi^0$ analysis, such the above cuts were not applied.

\subsection{$\pi^{0}$ reconstruction and selection}
There are three major steps in reconstructing a $\pi^0$ candidate in the FMS: cluster finding, shower shape fitting, and photon combination. The first step is to incorporate the adjacent towers with non-zero energy into clusters. A minimum energy threshold of 0.5~GeV  for 200~GeV  data and 1.0~GeV  for 500~GeV  data is applied to the reconstructed clusters to reject part of the charged hadron background. Due to the finite tower size, decay photons from a high energy $\pi^{0}$ tend to merge into one cluster, so the clusters need to be classified as one-photon-type or two-photon-type based on their size and energy distribution. After the clusters are found, a shower shape fitting procedure is applied to determine the energy and position of the photon candidate(s) for each cluster. An ideal shape of an electromagnetic shower is compared to the actual energy pattern of a cluster in the fitting. For a two-photon-type cluster, the separation between the two photons and their energy sharing are additional degrees of freedom that need to be determined. In the end, a list of photon candidates is generated and all pairs are used to build $\pi^0$ candidates.

Further $\pi^0$ selection includes a fiducial volume cut for the photons and other cuts for the $\pi^0$ candidates described below. The fiducial volume cut requires the photon position to be at least half of a tower width away from the outer and inner edge of the detector. For the $\pi^0$ reconstruction, there are further requirements as following:
\begin{itemize}
\item $p_\mathrm{T} > 2~\mathrm{GeV}/c,$
\item $2.7<\eta<4.0,$ 
\item $M_{\gamma\gamma}<0.3$~GeV/$c^2,$
\item $Z_{\gamma\gamma}=\left|\frac{E_1-E_2}{E_1+E_2}\right|<0.7$, where $E_1$ and $E_2$ are the energies of two photons.
\end{itemize}

\begin{figure}[htbp]
	\centering
	\includegraphics[scale = 0.3]{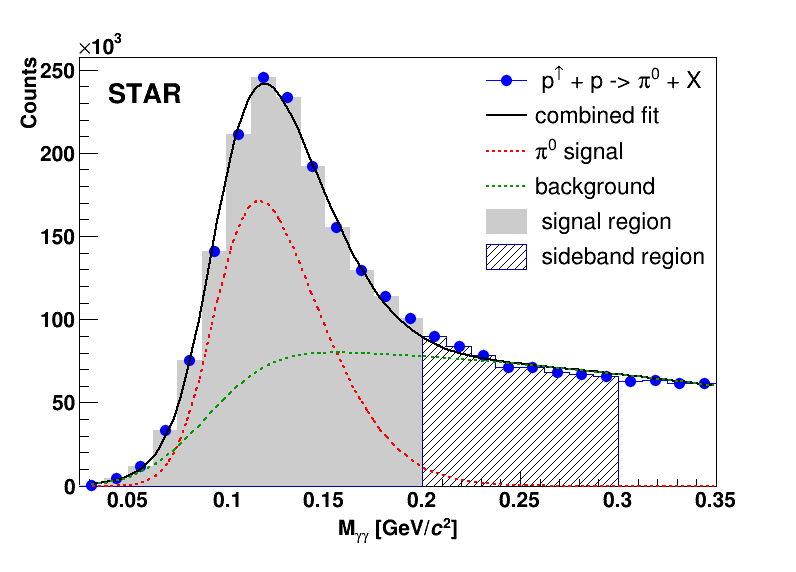}\\
	\caption{Example of invariant mass spectrum of the reconstructed gamma pairs in the FMS with an energy 38~GeV $< E_{\gamma\gamma}<$ 43~GeV in transversely polarized proton-proton collisions at $\sqrt s=$ 500~GeV. The mass spectrum is divided into signal region (0.0-0.2~GeV/$c^2$) and the sideband region (0.2-0.3~GeV/$c^2$). The dashed lines are the fit results for the $\pi^{0}$ signal and background. The solid line is the combined fit result. }\label{fg:pimass}	
\end{figure}

\Fgref{fg:pimass} shows an example of the invariant mass distribution of the reconstructed $\pi^0$ in 500~GeV  proton-proton collision data. The data were fitted to determine the signal fraction in the signal region (0.0-0.2~GeV/$c^2$) and sideband region (0.2-0.3~GeV/$c^2$). In this paper, skewed Gaussian functions in \Eqref{eq:skewG} are used to fit the signal and background shapes. The skewed Gaussian function has three shape parameters: the mean ($\xi$), the width ($\omega$) and the skewness ($\alpha$). The expected signal and background shape parameters for the two-photon invariant mass distribution in each $\pi^0$ energy bin are extracted from Monte Carlo (MC) simulation. The parameters of the skewed Gaussian functions are allowed to vary during fitting by 10-20\,\% depending on energy.  
The MC simulation used the standard STAR simulation framework based on GEANT 3~\cite{brun1987geant}, and PYTHIA 6.428~\cite{Sjostrand:2006za} as event generator with the CDF tune A~\cite{Field:2005sa}. 
\begin{equation}\label{eq:skewG}
\begin{split}
& f(x)=\frac{2}{\omega} \phi\left(\frac{x-\xi}{\omega}\right) \Phi\left(\alpha\cdot \frac{x-\xi}{\omega}\right)\\
& \phi(x)=\frac{1}{\sqrt{2 \pi}} e^{-\frac{x^{2}}{2}} \qquad \qquad\\
& \Phi(x)=\int_{-\infty}^{x} \phi(t) d t=\frac{1}{2}\left[1+\operatorname{erf}\left(\frac{x}{\sqrt{2}}\right)\right]
\end{split}
\end{equation}

For the $\pi^0$ in a jet, which is used in studying the Collins asymmetry, the $\pi^0$ reconstruction is slightly different and will be discussed in Sec.~II-E.2 in detail.

\subsection{Jet reconstruction}
In the measurement of the jet TSSA and the Collins asymmetry, the jet reconstruction is needed.  In this paper, the jet reconstruction is based on FMS energy deposits, and the anti-$k_\mathrm{T}$ algorithm is used within the FASTJET framework~\cite{fastjet}, with resolution parameter $R$ = 0.7. The photon candidates are used as basic building units in the jet reconstruction.
Similar as $\pi^0$ reconstruction, a minimum energy threshold of 0.5~GeV for 200~GeV data and 1.0~GeV  for 500~GeV data is applied to the photon candidates to reduce the possible charged hadron contribution.

The reconstructed jet energy is first corrected by subtracting the contributions from the underlying event, which is estimated utilizing the so-called ``off-axis cone method"~\cite{UE}. For a reconstructed jet, one first defines the axes of two cones at the same pseudo-rapidity as the reconstructed jet but at angles of $\pm\pi/2$ relative to the azimuthal angle of the jet. The cone parameter used is $R$ = 0.7. The energy density is calculated within each cone, where the jet area is given by the FASTJET package~\cite{fastjet} using the ghost particle technique. 

Then the jet kinematics are further corrected back to the ``particle level'', with a correction factor determined by a PYTHIA+GEANT simulation with same version and tune as in last subsection. We define the ``particle level" as the stable particles (photons here) produced in a proton-proton event in PYTHIA prior to the GEANT simulation of detector responses.  The correction factor ranges from 0.84 to 0.91 for 500 GeV data and from 0.97 to 1.03 for 200 GeV.  Note that the jet reconstructed this way is a partial jet in the sense that only photons are included, and will be referred to as an electromagnetic jet (EM-jet) in order to distinguish it from a full jet. Throughout the remainder of this paper, we will refer to an EM-jet as simply a jet, unless specified otherwise.

\begin{figure}[htbp]
	\centering
	\includegraphics[scale = 0.3]{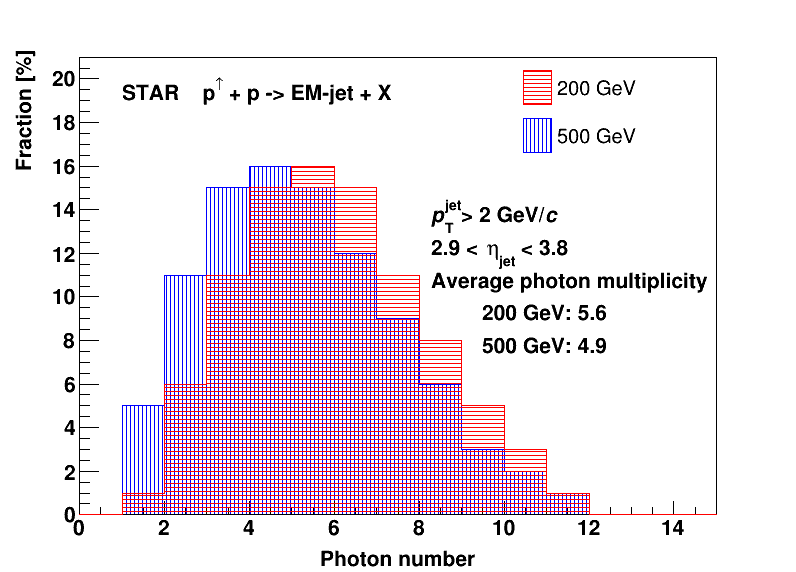}\\
	\caption{The observed EM-jet multiplicity distribution with the STAR FMS detector in transversely polarized proton-proton collisions 
	at 200 and 500~GeV.
	}\label{fg:jetinfo}
\end{figure}

In the jet reconstruction, no requirement on photon numbers is applied. \Fgref{fg:jetinfo} shows the measured photon multiplicity distribution for reconstructed jets with jet $p_\mathrm{T}$ greater than 2~GeV/$c$. The average photon multiplicity is 5.6 for 200~GeV data, and 4.9 for 500~GeV data. The higher photon energy cut at 500~GeV during jet reconstruction makes the observed multiplicity smaller than that at 200~GeV. 

\subsection{Asymmetry calculation}
\subsubsection{\textbf{E.1} $\pi^0$ and jet TSSA}

\EEqref{eq:asy0} shows the $\pi^0$ yield $N^{\uparrow}$ for spin ``up" of the $\pi^0$ production as a function of the azimuthal angle $\phi$ in transversely polarized proton-proton collisions. In this equation, $\epsilon $ stands for the efficiency of the detector, $\mathcal{L}$ for the beam luminosity, and $P$ for the beam polarization; the arrow indicates the spin direction of the beam. In order to eliminate effects due to a non-uniform detector efficiency and a time-dependent luminosity, the ``cross-ratio" method is used in calculating the asymmetry, see \Eqref{eq:asy8}. The ``cross-ratio" method~\cite{cross-ratio} takes advantage of the detector symmetry, which cancels efficiency and luminosity effects to leading order. In practice, $\phi$ is divided into 10 bins, which results in 5 data points on the r.h.s. of \Eqref{eq:asy8} as a function of $\cos\phi$, which are used to extract $A_\mathrm{N}^{\text{raw}}$. 
\begin{equation}\label{eq:asy0}
\begin{split}
N^{\uparrow}(\phi)& =\epsilon \mathcal{L}^{\uparrow} \sigma^{\uparrow} \\
& =\epsilon \mathcal{L}^{\uparrow} \left(1+P \cdot A_\mathrm{N} \cos \phi\right)\ \sigma_0
\end{split}
\end{equation}

\begin{equation}\label{eq:asy8}
P \cdot A_\mathrm{N}^{\text{raw}} \cos \phi= \frac{\sqrt{N^{\uparrow}(\phi) N^{\downarrow}(\phi+\pi)}-\sqrt{N^{\downarrow}(\phi) N^{\uparrow}(\phi+\pi)}}{\sqrt{N^{\uparrow}(\phi) N^{\downarrow}(\phi+\pi)}+\sqrt{N^{\downarrow}(\phi) N^{\uparrow}(\phi+\pi)}} 
\end{equation}
 
The raw asymmetry $A_\mathrm{N}^{\text{raw}}$ obtained using \Eqref{eq:asy8} has a contribution both from the signal and background. Assuming the background asymmetry $A_\mathrm{N}^{b k g}$ is constant over the mass region $0.0<M_{\gamma\gamma}<0.3$~GeV/$c^2$, the signal asymmetry $A_\mathrm{N}^{\pi^{0}}$ can be extracted by solving \Eqref{eq:ANcal0}. In these equations, the uncorrected signal ($A_\mathrm{N}^{\text {raw}_{sig}}$) and background ($ A_\mathrm{N}^{\text {raw}_{sb}}$) asymmetries are calculated  in the signal region $0.0<M_{\gamma\gamma}<0.2$~GeV/$c^2$ for the signal and the side-band region $0.2<M_{\gamma\gamma}<0.3$~GeV/$c^2$ for the background. The regions are shown in \fgref{fg:pimass}. The signal fractions in these two regions, $ f_{\text {sig}_{sig}}$ and $ f_{\text {sig}_{sb}}$, are obtained from fits to the $\pi^0$ invariant mass distribution as shown in \fgref{fg:pimass}. 
\begin{equation}\label{eq:ANcal0}
\begin{split}
&  A_\mathrm{N}^{\text {raw}_{sig}}=f_{\text {sig}_{sig}}A_\mathrm{N}^{\pi^{0}}+(1-f_{\text{sig}_{sig}})A_\mathrm{N}^{b k g} \\
&  A_\mathrm{N}^{\text {raw}_{sb}}=f_{\text {sig}_{sb}}  A_\mathrm{N}^{\pi^{0}}+\left(1-f_{s i g_{s b}}\right)  A_\mathrm{N}^{b k g}
\end{split}
\end{equation}

The extraction of the jet TSSA is almost the same as the $\pi^0$ TSSA using \Eqref{eq:asy8} except a slight difference on background part as detailed in Sec.III-C.

\subsubsection{\textbf{E.2} Collins asymmetry}
The extraction of Collins asymmetry ($A_\mathrm{UT}$) is  similar to that of the $\pi^0$ TSSA. Because of the different definition of the azimuthal angle, the cross-ratio method needs to be modified to account for the Collins angle $\phi_\mathrm{C}=\phi_\mathrm{S} - \phi_\mathrm{H}$, see \Eqref{eq:asy88}. For the Collins angles we follow the same definition as in Ref.~\cite{STARCollins}. $\phi_S$ is the angle between the upward spin direction of the polarized proton and the plane spanned by the momenta of the jet and the beam. The angle $\phi_\mathrm{H}$ is the angle between the jet-beam plane and the jet-pion plane determined by the $\pi^0$ momentum and the jet momentum. 
\begin{equation}\label{eq:asy88}
\begin{split}
P\cdot A_\mathrm{UT} & \sin \phi_\mathrm{C}= \\
& \frac{\sqrt{N^{\uparrow}(\phi_\mathrm{C}) N^{\downarrow}(\phi_\mathrm{C}+\pi)}-\sqrt{N^{\downarrow}(\phi_\mathrm{C}) N^{\uparrow}(\phi_\mathrm{C}+\pi)}}{\sqrt{N^{\uparrow}(\phi_\mathrm{C}) N^{\downarrow}(\phi_\mathrm{C}+\pi)}+\sqrt{N^{\downarrow}(\phi_\mathrm{C}) N^{\uparrow}(\phi_\mathrm{C}+\pi)}} 
\end{split}
\end{equation}

The $\pi^0$ reconstruction here is slightly different from the inclusive $\pi^0$ reconstruction as in Sec.II-C.  Since the $\pi^0$ is part of a jet, one needs to iterate over all combinations of photons within the jet. 
To avoid double-counting, photons can only be used once to reconstruct a $\pi^0$. In practice, the reconstruction starts with the highest energy $\pi^0$ candidate. If it passes all the selection cuts, its constituent photons will be excluded from the subsequent reconstruction. If it doesn't, the second highest energy $\pi^0$ candidate is checked, and so on, until a qualified candidate is found. The reconstruction continues with the next highest energy $\pi^0$ candidate from the remaining photons until all $\pi^0$ candidates have been evaluated. For this way of $\pi^0$ reconstruction, we do not perform a background subtraction for the Collins asymmetry. The possible influence from background is studied through the mass dependence of the asymmetry as discussed in Sec.~IV-D.

\section{Corrections and Systematic uncertainties}

\subsection{Energy uncertainty}
The photon energy uncertainty includes contributions from calibration, non-linear detector responses, and radiation damage. The contributions of the three types of energy uncertainties are 3.5\,\%, 1.5\,\%, and 2.2\,\% for 500 GeV data and 2.5\,\%, 1.5\,\%, and 0.5\,\% for 200 GeV data, respectively. The overall photon energy uncertainty is 4.4\,\% for 500 GeV data and 3.0\,\% for 200 GeV data.  

The $\pi^0$s and jets are composed of multiple photons. Their energy uncertainties are related to the energy of each of the constituent photons, and differ for every case. However, an upper limit for the energy uncertainty of a $\pi^0$/jet can be estimated using the constituent photon energy uncertainties. This estimation shows that the $\pi^0$ energy uncertainty is less than 4.4\,\% for 500 GeV data and 3.0\,\% for 200 GeV data. For the jet energy, additional uncertainties related with the energy correction factor to particle level are considered, which are estimated from simulation to be 6.4\,\% for 500~GeV and 8.0\,\% for 200~GeV data. These estimations show that the final jet energy uncertainty is less than 7.8\,\% for 500 GeV and 8.5\,\% for 200 GeV data. The details on these energy uncertainties can found be in Ref.~\cite{Zhanwen-thesis}

In this paper, the $\pi^0$ TSSA is extracted as a function of Feynman-$x$ and $p_\mathrm{T}$. Feynman-$x$ is defined as $x_\mathrm{F}=2 p_\mathrm{L} / \sqrt{s}$, and $p_\mathrm{L}$ is the longitudinal momentum. It approximately equals the $\pi^0$ energy divided by the proton beam energy. Its uncertainty is the same as the one for the $\pi^0$ energy. Since the photon angular uncertainty is much smaller than the energy uncertainty, the $\pi^0$ $p_\mathrm{T}$ uncertainty is also dominated by the $\pi^0$ energy. In summary, the uncertainties of $x_\mathrm{F}$ and $p_\mathrm{T}$ for $\pi^0$ TSSA are 4.4\,\% for 500~GeV and 3.0\,\% for 200~GeV data. The jet TSSA is presented versus $x_\mathrm{F}$ in Sec.~IV-C, and the $x_\mathrm{F}$ uncertainties are 7.8\,\% for 500~GeV and 8.5\,\% for 200~GeV data.

The Collins asymmetries are measured as a function of $z_\mathrm{em}$, which is the fraction of the $\pi^0$ energy over the jet energy, $z_\mathrm{em}=E_{\pi^0}/E_\mathrm{jet}$. The uncertainty of $z_\mathrm{em}$ can be estimated using the uncertainty on the ratio of $\pi^0$ energy and jet energy. This is found to be less than 8.9\,\% for 500~GeV and 9.0\,\% for 200~GeV data.

\subsection{The $\pi^0$ TSSA}
As discussed earlier, the two fractions $f_{\text{sig}_{sig}}$ and $f_{\text{sig}_{sb}}$ in \Eqref{eq:ANcal0} needed to calculate the TSSA, are obtained from fits to the $\pi^0$ invariant mass distribution. The uncertainty of the fractions as obtained from the fit are propagated to the $\pi^0$ TSSA as a source of systematic uncertainty. It is found that this uncertainty is up to 5.8\,\% of the magnitude of the asymmetry. This systematic uncertainty is smaller than the marker size in the TSSA result plots in the next section.

\subsection{The Jet TSSA}
The 200~GeV  data set contains a small number of jets reconstructed with energy far above the beam energy. These nonphysical events serve as a background under the jet signal, which may come from the pile up of non-collision background to normal events. The asymmetry of these events is consistent with zero. We assume these events  also exist at lower energy, which will decrease the measured jet TSSA. The asymmetries can be corrected using a background subtraction, with a correction factor $1/(1-r)$, where $r$ is the background fraction in the specified energy range. To estimate the background fraction, we choose the jet events in the energy range of 120~GeV  to 150~GeV  as pure background events. The energy spectrum of these events is found to be following a linear trend. A linear fit is done in this energy range and extrapolated to lower energy to estimate the background fraction. Results using this method show the highest background fraction is about 3\,\% for the highest $x_\mathrm{F}$ bin. 
 
\subsection{The Collins asymmetry}
The resolution of the Collins angle, $\phi_\mathrm{C}$, used in the calculation of the Collins asymmetry, is limited by the resolution of the photon position and jet axis. The resolution can be obtained from Monte Carlo simulations by comparing the reconstructed $\phi_\mathrm{C}$ on detector and particle levels. 
The smearing of this angle tends to underestimate the asymmetry and this effect can be corrected by multiplying a correction factor to the raw asymmetries. The resulting correction factor ranges from 1.01 to 1.04 in the region of $0.3 < z_\mathrm{em} < 0.9$.

\section{Results}
The clockwise-circulating RHIC beam (blue) faces the FMS. Single-spin asymmetries measured with respect to the blue beam polarization correspond to positive $x_\mathrm{F}$. The asymmetries with respect to the polarization of the counter-clockwise circulating beam (yellow), which corresponds to negative $x_\mathrm{F}$, are consistent with zero. This has been observed in multiple experiments~\cite{2008AN,PHENIX,Rhicf}. Therefore, the results with negative $x_\mathrm{F}$ are not shown.
Please note that there is a general scale uncertainty of 3.0/3.4\,\% for 200/500 GeV data from beam polarization for all spin asymmetries in this section, which is not included in the plots.

\subsection{The $\pi^0$ TSSA}

\begin{figure}[htbp]
	\centering
	\includegraphics[scale = 0.35]{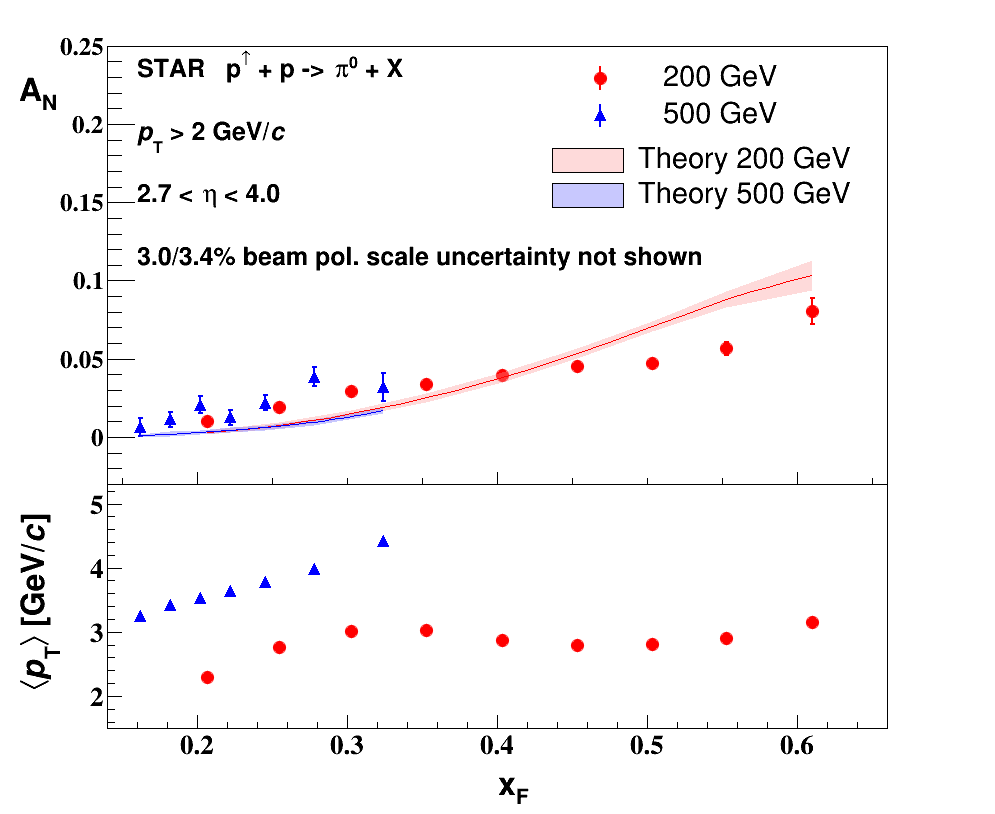}\\
	\caption{Transverse single-spin asymmetry ($A_\mathrm{N}$) as a function of $x_\mathrm{F}$ for $\pi^{0}$ production in transversely polarized proton-proton collisions at $\sqrt s=$ 200 and 500~GeV. The error bars are statistical uncertainties only. A systematic uncertainty up to 5.8\,\% of $A_N$ for each point is smaller than the size of the markers. The average $p_\mathrm{T}$ of the $\pi^{0}$ for each $x_\mathrm{F}$ bin is shown in the lower panel. Theory curves based on a recent global fit~\cite{Cammarota:2020qcw} are also shown.   }\label{fg:DrawAnComPt}
\end{figure}

\Fgref{fg:DrawAnComPt} shows the results of the $\pi^0$ TSSA for 200~GeV  (red points) and 500~GeV  (blue points) transversely polarized proton-proton collisions as a function of $x_\mathrm{F}$. The lower panel shows the average $\pi^0$ $p_\mathrm{T}$ for each  $x_\mathrm{F}$ bin.  The asymmetry increases with $x_\mathrm{F}$. The $x_\mathrm{F}$ of 200~GeV  data reach up to 0.6, where the largest asymmetry is observed. The results of both data sets are consistent in the overlapping region, $0.2<x_\mathrm{F}<0.35$. For both energies, the background asymmetries, which are not shown in the figure, are consistent with zero.

\begin{figure*}[htbp]		 			  
	\centering
	\includegraphics[scale = 0.3]{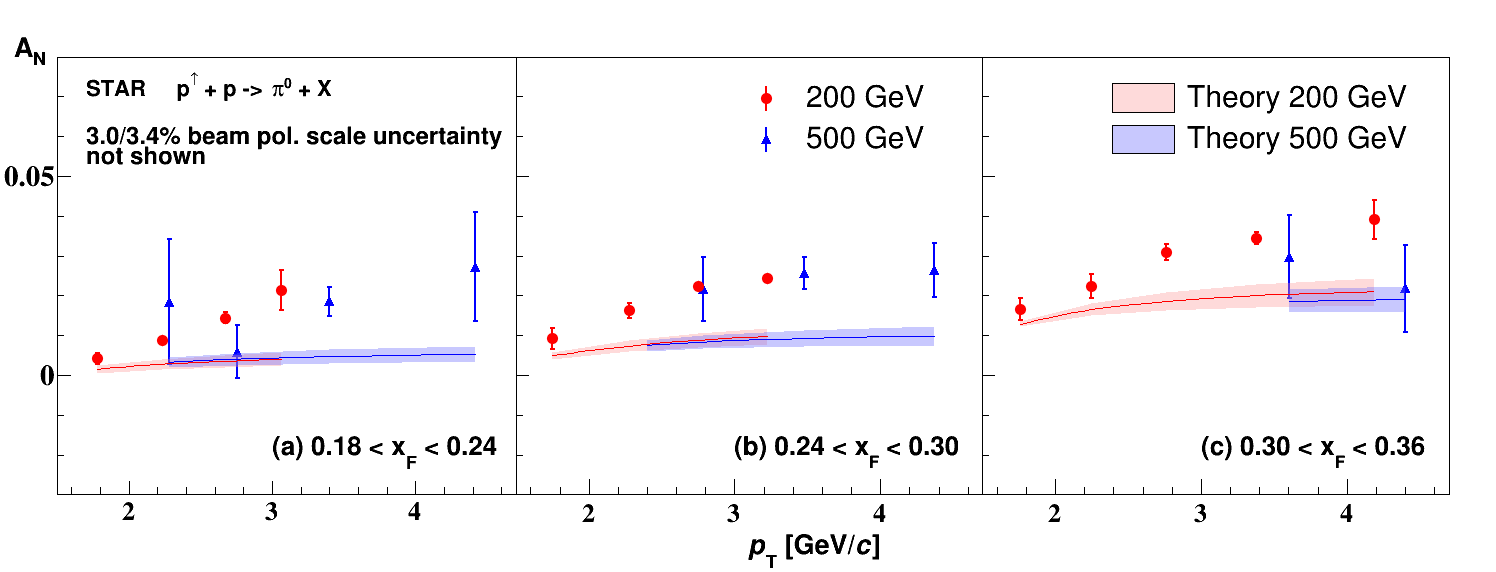}\\
	\caption{The transverse single-spin asymmetry as a function of the $\pi^{0}$ $p_\mathrm{T}$ for 
	three different $x_\mathrm{F}$ ranges (a)(b)(c) for transversely polarized proton-proton collisions at $\sqrt s=$ 200 and 500~GeV. The error bars are statistical uncertainties only. A systematic uncertainty up to 5.8\,\% of $A_N$ for each point is smaller than the size of the markers. Theory curves based on the recent global fit~\cite{Cammarota:2020qcw} are also shown.}
	\label{fg:DrawPt222AnCom}
\end{figure*}		

\Fgref{fg:DrawPt222AnCom} shows the TSSA result as a function of $\pi^0$ $p_\mathrm{T}$, in the overlap $x_\mathrm{F}$ region, $0.18 < x_\mathrm{F} < 0.36$, for the two data sets. The three panels represent different regions in $x_\mathrm{F}$. Although the statistics for the 500~GeV  data are limited, it can be seen that the results at the two beam energies are consistent. In the $x_\mathrm{F}$ regions covered by the data, the 200~GeV  results show the asymmetries rise with $p_\mathrm{T}$, clearly indicating a dependence of the asymmetry on $p_\mathrm{T}$ and $x_\mathrm{F}$. This is consistent with similar observations in previous STAR measurements~\cite{2008AN}. More details regarding the $p_\mathrm{T}$ dependence at 200~GeV  can be found in Ref.~\cite{nucl_dep_paper}.

\begin{figure}[htbp]
	\centering
	\includegraphics[scale = 0.35]{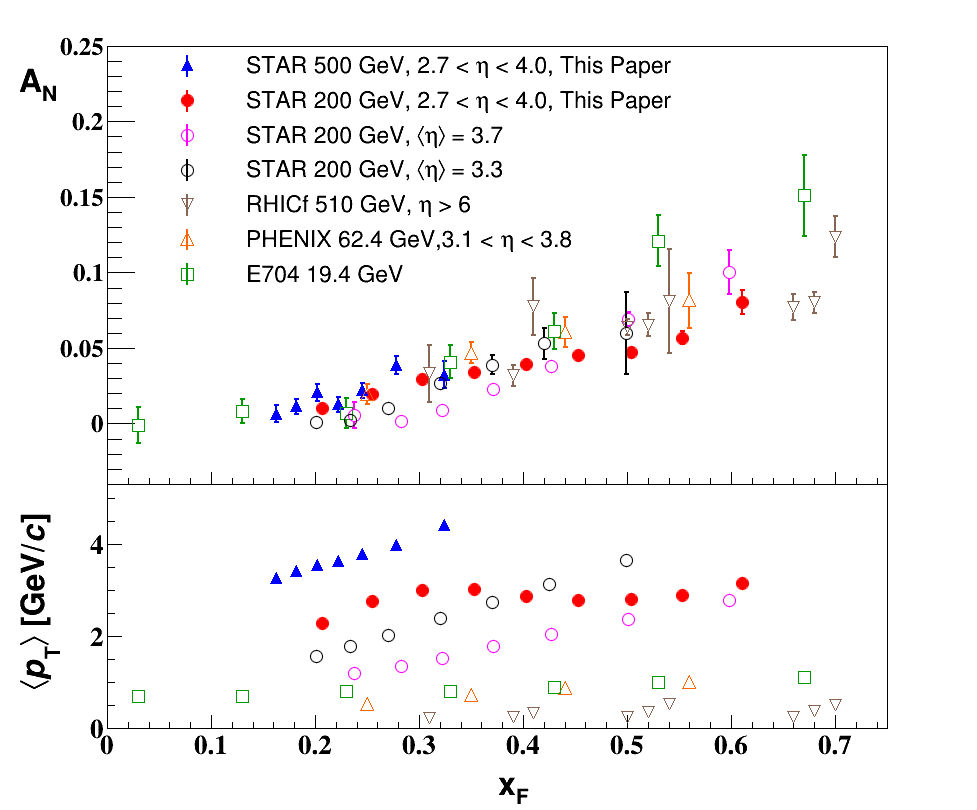}\\
	\caption {Comparison of this measurement of the transverse single-spin asymmetry as a function of $x_\mathrm{F}$ for inclusive $\pi^{0}$ with previous measurements from $\sqrt s=$ 19.4~GeV  to $\sqrt s=$ 510~GeV  in transversely polarized proton-proton collisions. The error bars are statistical uncertainties only. The average $p_\mathrm{T}$ of the $\pi^{0}$ for each $x_\mathrm{F}$ bin is shown in the lower panel.}
	\label{fg:ExpALL}
\end{figure}

\Fgref{fg:ExpALL} shows the comparison of these STAR  results with the other existing measurements in transversely polarized proton-proton collisions. They include previous STAR measurements using the FPD detector~\cite{2008AN}, results from the RHICf experiment~\cite{Rhicf}, the PHENIX experiment~\cite{PHENIX}, and the E704 experiment~\cite{E7042} at Fermi National Accelerator Laboratory. The average $p_\mathrm{T}$ of the $\pi^{0}$ for each $x_\mathrm{F}$ bin is shown in the lower panel. The $\pi^0$ TSSA results in this paper are consistent with the other measurements. This can only be explained with a very weak scale dependence of the $\pi^0$ TSSA for a $\sqrt s$ range of 19.4 to 510~GeV. The earlier 200~GeV  STAR results~\cite{2008AN} seem to be slightly lower than the current 2015 results in the range of $x_\mathrm{F}<0.4$. This could be explained by the $p_\mathrm{T}$ dependence of the TSSA results. From the above discussion, the TSSA results are not only a function of $x_\mathrm{F}$, but also a function of the $p_\mathrm{T}$. At the same $x_\mathrm{F}$ range, the asymmetries rise with the $p_\mathrm{T}$ in the region 
$1~\mathrm{GeV}/c < p_\mathrm{T} < 3$~GeV/$c$. The lower panel of \fgref{fg:ExpALL} shows that the mean $p_\mathrm{T}$ as a function of $x_\mathrm{F}$ in the region of $x_\mathrm{F}<0.4$ in this paper are higher than those of the earlier 200~GeV  STAR results.

\subsection{The TSSA for isolated $\pi^0$}
In searching for the origin of the transverse single-spin asymmetry, one particularly interesting aspect is the topological dependence of $\pi^0$ TSSAs, meaning one divides the $\pi^0$ sample into sub-groups based on the event structure. One group contains the isolated $\pi^0$s, which refers to the $\pi^0$s with no other surrounding photons. The other group contains the non-isolated $\pi^0$s, which are accompanied by other photons. In practice, the energy fraction $z_\mathrm{em}$, which is the $\pi^0$ energy over the jet energy, is used to determine whether or not a $\pi^0$ is isolated. Two photons alone can be reconstructed as a jet, so a $\pi^0$ would be identified as isolated when its $z_\mathrm{em}$ is close to 1. In the following step, one applies $z_\mathrm{em}>0.98$ to select isolated  $\pi^0$ and $z_\mathrm{em}<0.9$ for the non-isolated ones. The gap ensures a clean separation between the two groups.

In this way, both types of $\pi^0$s always correlate with a jet. Therefore, its constituent photons should be limited within the same jet. The $\pi^0$ selection and asymmetry calculation remain the same. The jet resolution  parameter $R = 0.7$ indicates the area where the $\pi^0$ is considered to be isolated.  
\begin{figure}[htbp]
	\centering
	\includegraphics[scale = 0.35]{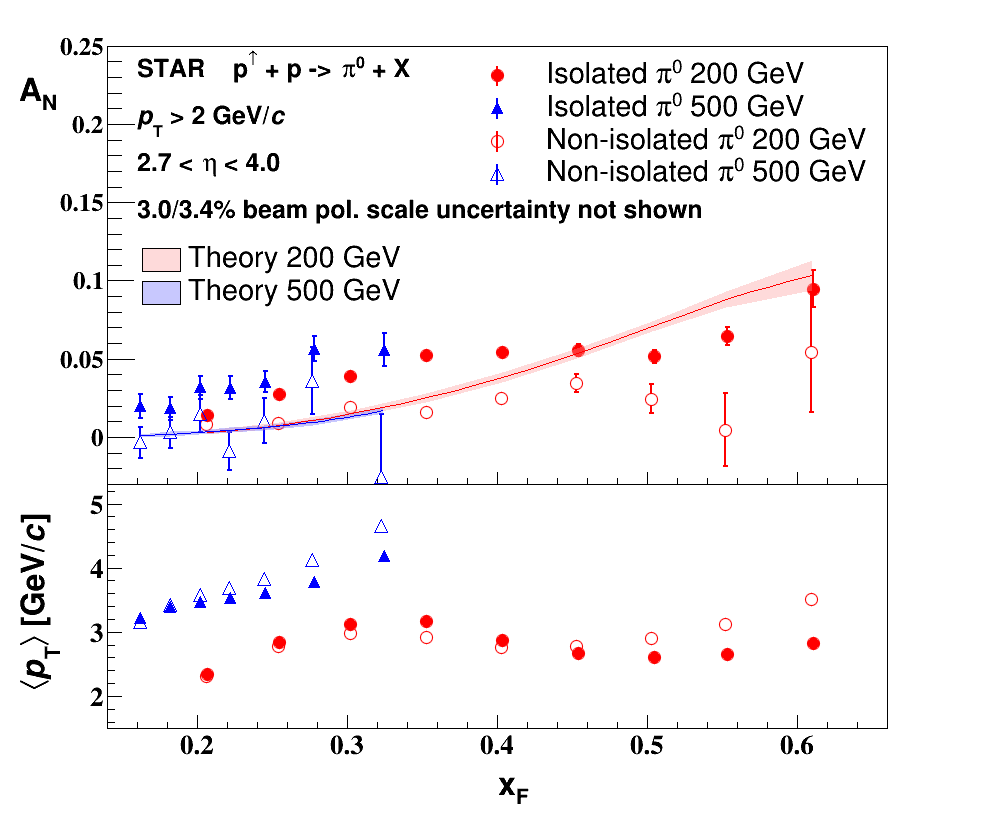}\\
	\caption{The transverse single-spin asymmetry as a function of $x_\mathrm{F}$ for the isolated and non-isolated $\pi^{0}$ in transversely polarized proton-proton collisions at $\sqrt s=$ 200 and 500~GeV. The error bars are statistical uncertainties only. A systematic uncertainty up to 5.8\,\% of $A_N$ for each point is smaller than the size of the markers. Theory curves based on a recent global fit~\cite{Cammarota:2020qcw} are also shown. The average $p_\mathrm{T}$ of the $\pi^{0}$ for each $x_\mathrm{F}$ bin is shown in the lower panel. }\label{fg:DrawIsoComPt}
\end{figure}

\Fgref{fg:DrawIsoComPt} shows the TSSA of these two types of $\pi^{0}$. Although the asymmetries of both types increase with $x_\mathrm{F}$, their magnitudes are significantly different. The asymmetries for the isolated $\pi^0$ are clearly larger than the asymmetries for the non-isolated $\pi^0$. 
This result suggests there could be different mechanisms in play to explain the large asymmetries shown in \fgref{fg:DrawAnComPt}. The non-isolated $\pi^0$s are considered to be part of a jet, which has fragmented from a parton, while the underlying subprocess for the isolated ones is not yet clear. One possible explanation is that a significant part of the isolated $\pi^0$s are from diffractive processes~\cite{RHICcoldQCD}, which needs further confirmation. The theoretical descriptions mentioned in the introduction would mainly apply to the TSSA of the non-isolated $\pi^0$s, which usually assume all the $\pi^0$s come from parton fragmentation, for example in a recent global analysis~\cite{Cammarota:2020qcw}.
A recent measurement of TSSA for very forward $\pi^0$ in transversely polarized proton-proton collisions by the RHICf experiment also indicates that the diffractive process could give a sizable asymmetry~\cite{Rhicf}.

\begin{figure}[htbp]
	\centering
	\includegraphics[scale = 0.32]{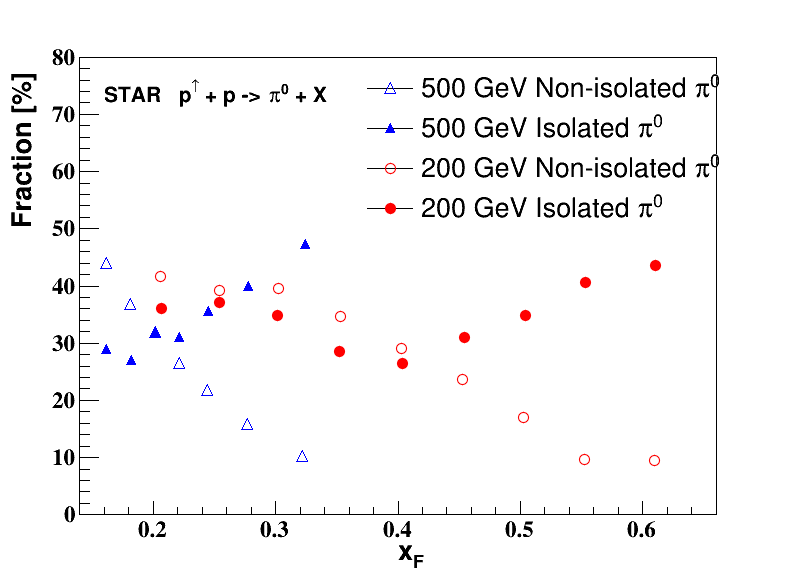}\\
	\caption{Fractions of isolated and non-isolated $\pi^{0}$ to the overall inclusive $\pi^{0}$ sample in the mass region 0-0.3~GeV/$c^2$, after background subtraction. The missing fraction mainly includes the events between the isolated cuts: $0.9<z_\mathrm{em}<0.98$. }\label{fg:frac}
\end{figure}

To understand the contributions from isolated and non-isolated $\pi^0$ to the overall $\pi^0$ TSSA, Fig.~\ref{fg:frac} shows the fractions of each type in the overall $\pi^0$ sample. It is noted that these fractions are background corrected to ensure the  fractions represent the $\pi^0$ signal only. It can be seen that, for each data set, the isolated $\pi^0$ plays an important role in the high $x_\mathrm{F}$ region where the asymmetry is significantly larger. In Ref.~\cite{nucl_dep_paper}, a somewhat different isolation criterion was used, but the same conclusion was obtained that the isolated $\pi^0$ have larger TSSA than the non-isolated $\pi^0$ in $p$+Al and $p$+Au collisions in addition to proton-proton collisions.

\subsection{The Jet TSSA}
\begin{figure}[htbp]
	\centering
	\includegraphics[scale = 0.35]{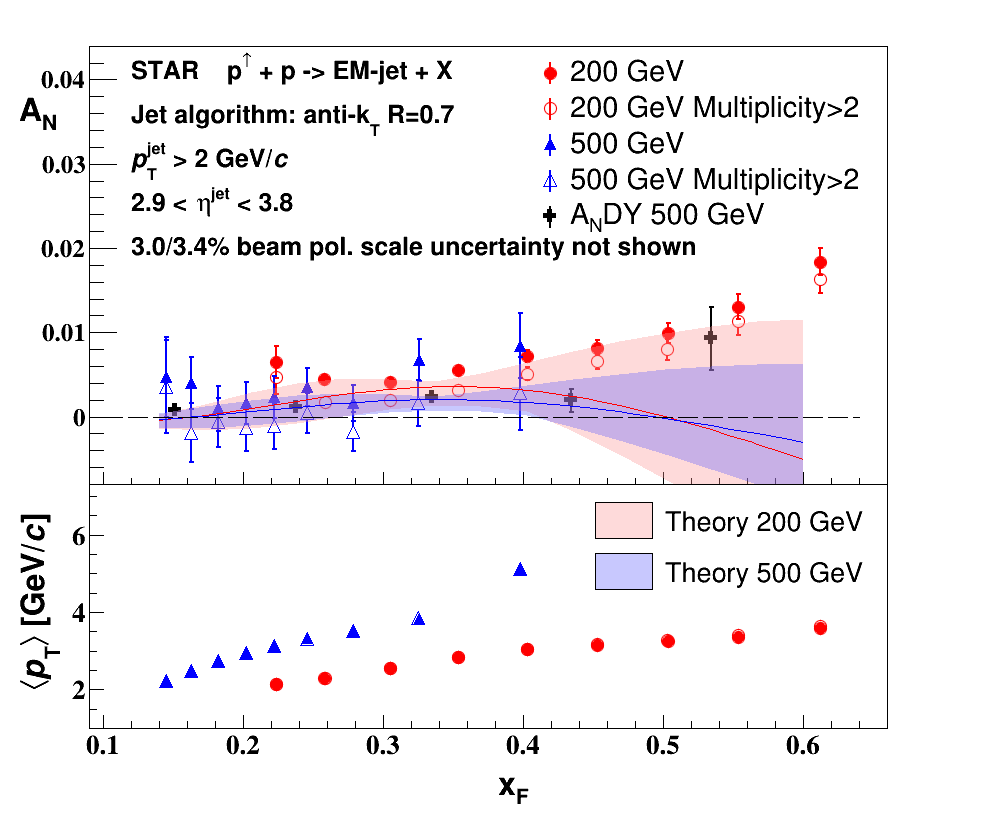}\\
	\caption{Transverse single-spin asymmetry as a function of $x_\mathrm{F}$ for electromagnetic jets in transversely polarized proton-proton collisions at $\sqrt s=$ 200 and 500~GeV. The error bars are statistical uncertainties only and the systematic uncertainties are negligible. The results that require more than two photons observed inside a jet are shown as open symbols. The previous measurements for full jets at 500~GeV  reported by the $A_\mathrm{N}DY$ Collaboration~\cite{ANDY} are  also plotted. Theory curves~\cite{jetAN2} for TSSA of full jets at mean rapidity $\langle y\rangle$ = 3.25 for 200~GeV~(red)  and $\langle y\rangle$ = 3.57 for 500~GeV~(blue)   are also shown.  
	 The average $p_\mathrm{T}$ of the jet for each $x_\mathrm{F}$ bin is shown in the lower panel.}\label{fg:jetAn}
\end{figure}

\Fgref{fg:jetAn} shows the results of the jet TSSA as a function of $x_\mathrm{F}$ for both data sets. The solid symbols in the figure represent the results that have no limitation on the photon multiplicity when reconstructing the jet, while the open symbols represent the results that required the observed photon multiplicity in the jet to be greater than 2. 
The asymmetries are non-zero and increase with $x_\mathrm{F}$, similar to the $\pi^0$ TSSA. The consistency of the 200 and 500~GeV  jet asymmetries in the overlap region suggests a weak energy dependence. However, the jet asymmetries are much smaller than the $\pi^0$ ones in \fgref{fg:DrawAnComPt} for the same $x_\mathrm{F}$. Theoretically, the jet asymmetry is believed to be dominated by initial-state effects related with the Sivers function.

Since a single photon or two photons can be reconstructed as a jet, the isolated $\pi^0$ sample described earlier is part of the jet sample and therefore enhances the overall jet TSSA. The open symbols in \fgref{fg:jetAn} show the TSSA for jets with a measured photon multiplicity greater than 2. The jet TSSAs with a minimum multiplicity requirement are smaller than the ones without this requirement, while the $p_\mathrm{T}$ at each $x_\mathrm{F}$ of the two samples is almost the same. The 200~GeV results are significantly larger than zero, while the 500~GeV results are consistent with zero within uncertainties, which may indicate a stronger energy dependence than what was observed for the $\pi^0$ TSSA.  

The black crosses in \fgref{fg:jetAn} represent the results from the $A_\mathrm{N}DY$ Collaboration at RHIC~\cite{ANDY} with transversely polarized proton-proton collisions at 500~GeV. The $A_\mathrm{N}DY$ experiment measured jets using an electromagnetic and a hadronic calorimeter to reconstruct both the electromagnetic and hadronic components of jets. The $A_\mathrm{N}DY$ result suggests the jet TSSA are very small and they are close to the STAR jet TSSA result measured at 500~GeV  with the minimum multiplicity requirement. The consistency of these two results suggests that the TSSA for EM-jets probes the same underlying physics as full jets.

\subsection{The Collins asymmetry}

The Collins effect is defined as a non-uniform azimuthal distribution of a particle's $p_\mathrm{T}$ in the hadronization of a transversely polarized quark~\cite{Collins:1992kk}. By measuring the Collins asymmetry of $\pi^0$ within a jet, one can directly study the fragmentation process contribution to the single-spin asymmetry at forward rapidities. The Collins angle ($\phi_\mathrm{C}$) in \Eqref{eq:asy88} is defined in the same way as in Ref.~\cite{STARCollins}. 
The resolution of the Collins angle is the major source of the asymmetry uncertainty. If the direction of the $\pi^0$ momentum is close to the jet thrust axis, for example at high $z_\mathrm{em}$, the uncertainty of the $\phi_\mathrm{C}$ angle becomes large. Therefore, a $\Delta R$ cut, $\Delta R=\sqrt{((\eta_{\pi^0}- \eta_\mathrm{jet})^2+(\phi_{\pi^0}-\phi_\mathrm{jet})^2}$, has been applied in the analysis to reject such events. The value of this cut was balanced between the benefit of excluding those events with large uncertainty and the loss of statistics at high $z_\mathrm{em}$. We determined $\Delta R > 0.04$ to be the best choice, which is the same as in Ref.~\cite{STARCollins}. 

As mentioned in Sec.~II-E.2, there is no background subtraction for the Collins asymmetry. Nevertheless, the influence of possible background can be studied through the mass dependence of the asymmetry. The $\pi^0$ signal is concentrated in the mass region $M_{\gamma\gamma}<$ 0.2~GeV/$c^2$, whereas the background fraction changes significantly as a function of mass from the region $M_{\gamma\gamma}<$ 0.2~GeV/$c^2$ to the region $M_{\gamma\gamma}>$ 0.2~GeV/$c^2$. A comparison of the Collins results in the region of (0, 0.2~GeV/$c^2$) and those in the region of (0.2, 0.3~GeV/$c^2$) did not show a clear mass dependence in both data sets.

\begin{figure}[htbp]
	\centering
	\includegraphics[scale = 0.32]{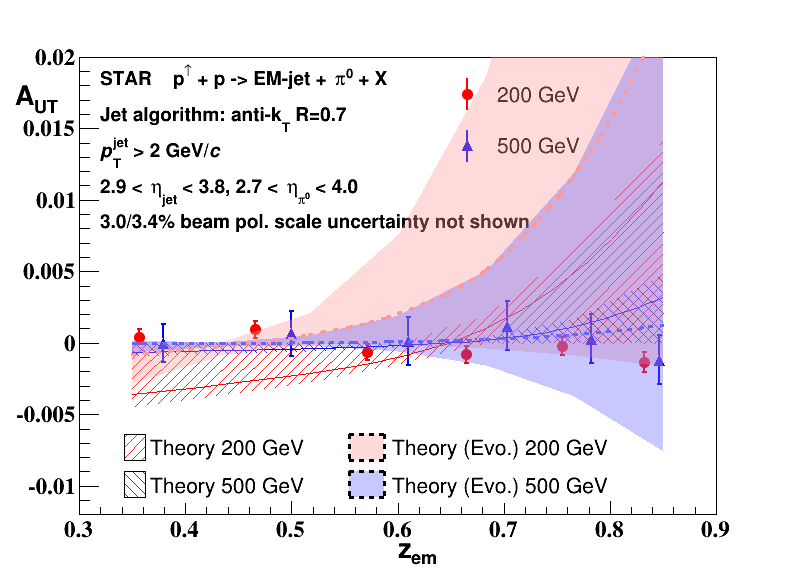}\\
	\caption{The Collins asymmetry for $\pi^{0}$ in an electromagnetic jet for transversely polarized proton-proton collisions at $\sqrt s=$ 200 and 500~GeV. The error bars are statistical uncertainties only and the systematic uncertainties are negligible. Theory curves for the Collins asymmetry of a $\pi^0$ in a full jet with or without TMD evolution~\cite{Kang:2017btw} are also shown. }\label{fg:DrawCollCom}
\end{figure}

The jet $p_\mathrm{T}$ is required to be larger than 2~GeV/$c$. The average jet $p_\mathrm{T}$ is 3.8~GeV/$c$ for 500 GeV data and 3.0~GeV/$c$ for 200 GeV data. The average jet pseudorapidity is 3.1 for 500 GeV data and 3.3 for 200 GeV data. \Fgref{fg:DrawCollCom} shows the measured Collins asymmetries ($A_{UT}$) originating from the final-state effect, for both the 200 and 500~GeV  data. Both results show very small asymmetries within uncertainties. 

\begin{figure}[htbp]
    \centering
	    \includegraphics[scale = 0.32]{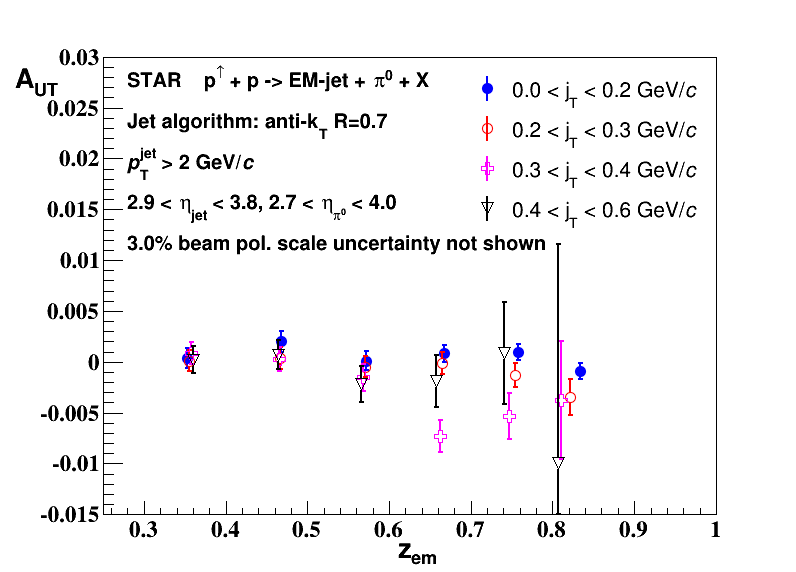}\\
		\caption{The $j_\mathrm{T}$ dependence of the $\pi^{0}$ Collins asymmetry in transversely polarized proton-proton collisions at $\sqrt s=200$~GeV. The error bars are statistical uncertainties only and the systematic uncertainties are negligible. }\label{fg:DrawCollRun15JtALL} 
 \end{figure}

The $\pi^0$ momentum transverse to the jet axis, $j_\mathrm{T}$, can be used to measure how close the $\pi^0$ is to the jet axis. An investigation of the dependence of the Collins asymmetry on $j_\mathrm{T}$ at 200~GeV  is presented in \fgref{fg:DrawCollRun15JtALL}. The Collins asymmetries are separated into four $j_\mathrm{T}$ bins. It is found that the asymmetries for $j_\mathrm{T}>0.2$~GeV/$c$ show a tendency to be negative. This $j_\mathrm{T}$ dependence can be used to further constrain TMD models.  


\subsection{Comparison to models}
We compare our results to the theoretical calculations that can be seen in Figs.~\ref{fg:DrawAnComPt},~\ref{fg:DrawPt222AnCom},~\ref{fg:DrawIsoComPt},~\ref{fg:jetAn}, and \ref{fg:DrawCollCom}. The calculations of $\pi^0$ TSSA~\cite{Cammarota:2020qcw}, jet TSSA~\cite{jetAN2} and Collins asymmetry~\cite{Kang:2017btw} are based on the TMD and Collinear Twist-3 functions that have been extracted from semi-inclusive deep inelastic scattering, Drell-Yan, $e^+e^{-}$ annihilation into hadron pairs, and transversely polarized proton-proton collisions that included also previous forward $\pi^0$ and charged hadron TSSA data from RHIC.
The calculations refer to the kinematics of the data in this paper to account for the known kinematic dependencies of the measurements.

As shown in Figs.~\ref{fg:DrawAnComPt} and \ref{fg:DrawPt222AnCom}, the calculations have almost no energy dependence~\cite{Cammarota:2020qcw}. They underestimate the $\pi^0$ TSSA in the lower $x_\mathrm{F}$ region for both 200 and 500~GeV  data, but overestimate it in the higher $x_\mathrm{F}$ region where 200~GeV  data are available. In \fgref{fg:DrawPt222AnCom}, the calculations show the same trend of asymmetries rising with $p_\mathrm{T}$ as the data, but the magnitude of the predicted asymmetry is much smaller than the measurements. 

The theory curves in \fgref{fg:DrawIsoComPt} are identical to the ones in \fgref{fg:DrawAnComPt}. In the $x_F$ region lower than 0.3, they can describe the non-isolated $\pi^0$ TSSA measurements, in which these $\pi^0$s are considered to originate from fragmentation. The theory curve in this region is mostly constrained by the Sivers/Collins inputs from SIDIS data~\cite{Cammarota:2020qcw}.

For the jet TSSA in \fgref{fg:jetAn}, the calculation for 500~GeV  is consistent with the measurement that has the minimum photon multiplicity requirement and also the full jet result from the $A_\mathrm{N}DY$ experiment. However, the calculation for 200~GeV  predicts the asymmetry to fall with $x_\mathrm{F}$, which contradicts our measurement. It is noted that the theoretical uncertainty bands are substantial~\cite{jetAN2}. 

For the Collins asymmetry in \fgref{fg:DrawCollCom}, two sets of theory curves represent the cases with or without the TMD evolution being taken into account~\cite{Kang:2017btw}. Our $j_\mathrm{T}$ combined results for both collision energies are consistent with zero, which are within the uncertainty bands of the two calculations.

\section{Conclusion}

We report the measurements of transverse single-spin asymmetries for $\pi^0$s in the forward rapidity region in transversely polarized proton-proton collisions at 200~GeV and 500~GeV using the FMS detector at STAR. The measurement at 200~GeV was done with the largest data sample thus far. The asymmetries increase with $x_\mathrm{F}$. No energy dependence was found when comparing the current results with previous data at RHIC and FNAL with center-of mass energies as low as 19.4~GeV. 
The transverse single-spin asymmetries for isolated and non-isolated $\pi^0$ at both 200~GeV  and 500~GeV  were also presented. The asymmetries of isolated $\pi^0$s are significantly larger than those of non-isolated $\pi^0$s. 

The transverse single-spin asymmetries for electromagnetic jets were measured with the FMS in transversely polarized proton-proton collisions at both 200~GeV and 500~GeV. The 500~GeV result with a minimum photon multiplicity requirement is consistent with zero, which coincides with the full jet measurement from the $A_\mathrm{N}DY$ experiment. The 200~GeV results are small, but clearly non-zero within uncertainties. 

Collins asymmetries for $\pi^0$s within an electromagnetic jet were measured in transversely polarized proton-proton collisions at both 200~GeV and 500~GeV. The asymmetries are small across the $z_\mathrm{em}$ bins and might exhibit a $j_\mathrm{T}$ dependence at 200~GeV. The latter could help to constrain TMD models and needs theoretical predictions. 

These new data provide important information for understanding the underlying physics mechanism for the transverse single-spin asymmetry. 
In particular, the observed small TSSA for non-isolated $\pi^0$s and also small Collins asymmetries with EM jets suggest that the Collins effect itself cannot account for the observed $\pi^0$ TSSA.
On the other hand, the observed small TSSA for electromagnetic jets indicates the contribution from the Sivers effect cannot be the dominant source of $\pi^0$ TSSA, either. 
The sizable TSSA for isolated $\pi^0$ thus indicates a new mechanism, likely diffractive process, could be a significant source for the $\pi^0$ TSSA in transversely polarized proton-proton collisions at RHIC, and more theory efforts and dedicated measurements are called for to have a complete understanding on this aspect.

\section*{Acknowledgement}
We thank the RHIC Operations Group and RCF at BNL, the NERSC Center at LBNL, and the Open Science Grid consortium for providing resources and support.  This work was supported in part by the Office of Nuclear Physics within the U.S. DOE Office of Science, the U.S. National Science Foundation, the Ministry of Education and Science of the Russian Federation, National Natural Science Foundation of China, Chinese Academy of Science, the Ministry of Science and Technology of China and the Chinese Ministry of Education, the Higher Education Sprout Project by Ministry of Education at NCKU, the National Research Foundation of Korea, Czech Science Foundation and Ministry of Education, Youth and Sports of the Czech Republic, Hungarian National Research, Development and Innovation Office, New National Excellency Programme of the Hungarian Ministry of Human Capacities, Department of Atomic Energy and Department of Science and Technology of the Government of India, the National Science Centre of Poland, the Ministry  of Science, Education and Sports of the Republic of Croatia, RosAtom of Russia and German Bundesministerium fur Bildung, Wissenschaft, Forschung and Technologie (BMBF), Helmholtz Association, Ministry of Education, Culture, Sports, Science, and Technology (MEXT) and Japan Society for the Promotion of Science (JSPS).

\bibliography{paper}

\end{document}